\documentclass[pra,twocolumn]{article}
\usepackage{amsmath,amsfonts,amssymb,amsthm}
\usepackage{eqnarray}
\usepackage{graphics,graphicx}
\usepackage{mathcomp} 
\usepackage{srcltx}
\usepackage{tabularx,ragged2e,booktabs}
\usepackage{ulem} 
\usepackage{xcolor}

\usepackage{authblk}
\usepackage{enumitem} 
\setlist[description]{leftmargin=0.0cm,labelindent=0.0cm}
\usepackage{mathtools, cuted}
\usepackage[section]{placeins}
\usepackage[superscript]{cite} 
\let\OLDthebibliography\thebibliography
\renewcommand\thebibliography[1]{
  \OLDthebibliography{#1}
  \setlength{\parskip}{0pt}
  \setlength{\itemsep}{0pt plus 0.3ex}
}

\newcommand{\beq}{\begin{equation}}
\newcommand{\eeq}{\end{equation}}
\newcommand{\bea}{\begin{eqnarray}}
\newcommand{\eea}{\end{eqnarray}}
\newcommand{\ket} [1] {|#1\rangle}
\newcommand{\bra} [1] {\langle#1|}

\def\({\left(}
\def\){\right)}
\def\[{\left[}
\def\]{\right]}

\title{\textbf{Overcoming the rate-distance barrier of quantum key distribution without using quantum repeaters}}

\author[1]{M. Lucamarini}
\author[1]{Z.~L.~Yuan}
\author[1]{J.~F.~Dynes}
\author[1]{A.~J.~Shields}
\affil[1]{Toshiba Research Europe Ltd, 208 Cambridge Science Park, Cambridge CB4 0GZ, United Kingdom}
\date{}
\begin{document}
\maketitle



\noindent \textbf{Quantum key distribution (QKD)\cite{BB14,Eke91} allows two distant parties to share encryption keys with security based on physical laws. Experimentally, it has been implemented with optical means, achieving key rates of 1.26~Megabit/s over 50~kilometres (km) of standard optical fibre\cite{CFL+14} and 1.16 bit/hour over 404~km of ultralow-loss fibre in a measurement-device-independent configuration\cite{YCY+16}. Increasing the bit rate and range of QKD is a formidable, but important, challenge. A related target, currently considered unfeasible without quantum repeaters\cite{BDC+98,DLC+01,SSD+11}, is overcoming the fundamental rate-distance limit of point-to-point QKD\cite{TGW14,PLO+17}. 
Here we introduce a conceptually new scheme where pairs of phase-randomised optical fields are first generated at two distant locations and then combined at a central measuring station. The fields imparted with the same random phase are ``twins'' and can be employed to distill a quantum key, as we prove under an explicit security assumption. The key rate of this twin-field QKD (TF-QKD) shows the same dependence on distance as a quantum repeater, scaling with the square-root of the channel transmittance irrespective of whom is in control of the measuring station. Differently from a quantum repeater, however, the new scheme is feasible with current technology and presents manageable levels of noise even on 550~km of standard optical fibre. This is promising to overcome the QKD rate-distance barrier and to greatly extend the range of secure quantum communications.}

\begin{figure*}
\centering
\includegraphics[width=0.75\textwidth]{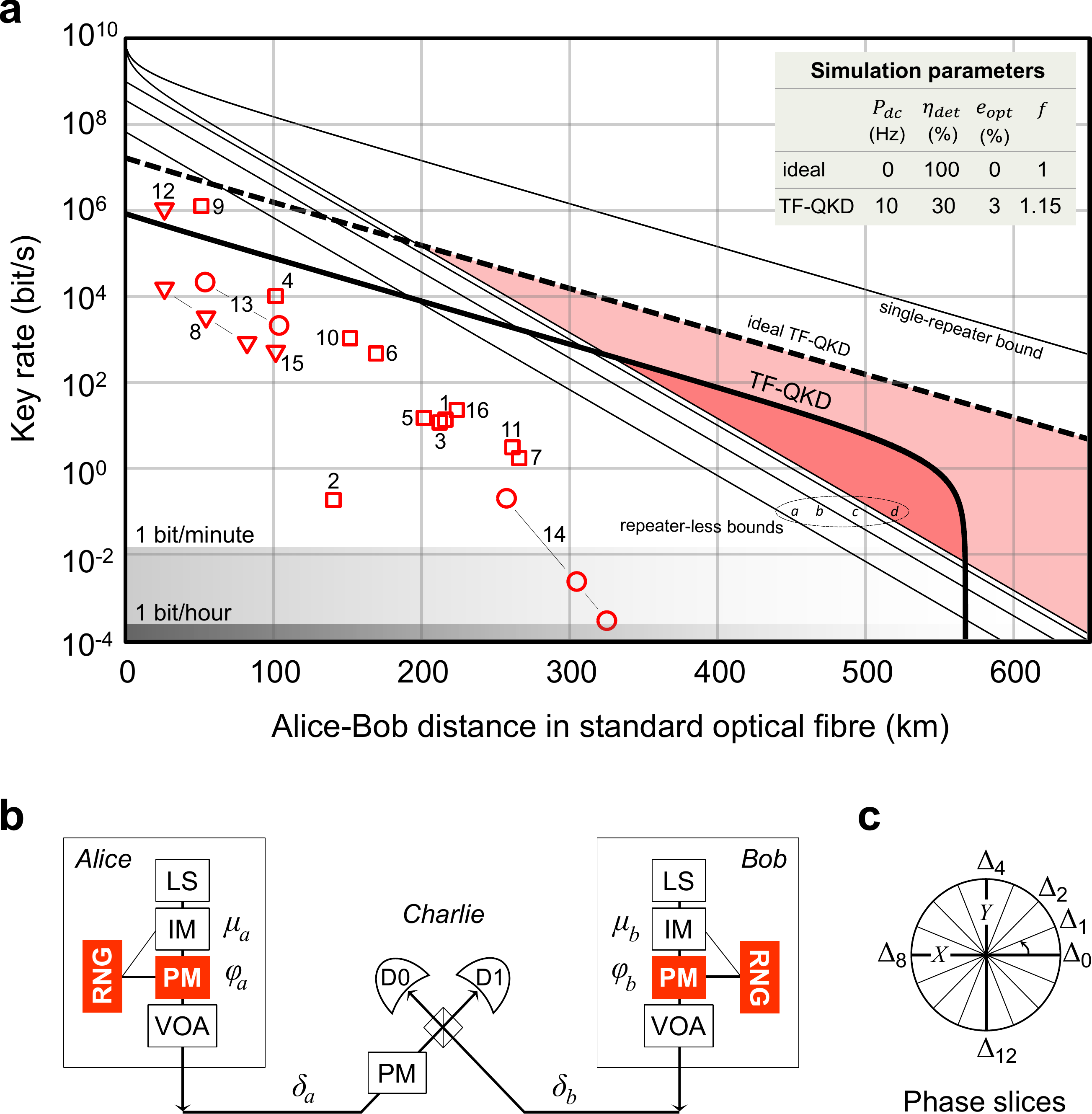}
\caption{\textbf{Scheme to overcome the rate-distance limit of QKD}.
\textbf{a}, Theoretical bounds (lines) and experimental results (symbols) for fibre-based quantum schemes (details in SI). To make a homogeneous comparison, all the distances have been normalised to the length $L$ of a standard optical fibre with attenuation coefficient $\alpha=0.2$~dB/km. Letter-code for the theoretical bounds are: $a$, decoy-state MDI-QKD; $b$, decoy-state QKD; $c$, single-photon QKD; $d$, secret key capacity\cite{PLO+17}. The single-repeater bound is from Ref.~[\citen{Pir16}]. Symbol-code for the experimental results: squares, triangles and circles are for QKD, continuous-variable QKD and MDI-QKD, respectively.
TF-QKD is the scheme described in this work. The solid (dashed) line is for the realistic (ideal) TF-QKD key rate given in Eq.~\eqref{TF-QKD-rate}. \underline{Inset}: Parameters used for numerical simulations. $P_{dc}$, dark count probability; $\eta_{det}$, total detection efficiency; $e_{opt}$, channel optical error rate; $f$, error correction coefficient.
\textbf{b}, Setup to implement TF-QKD. The light sources (LS) generate pulses whose intensities $\mu_{a,b}$ are randomly varied by the intensity modulators (IM) to implement the decoy-state technique\cite{Hwa03,Wan05,LMC05}. Phase modulators (PM) are combined with random number generators (RNG) to encode each light pulse with phases $\varphi_{a,b}$, which include the random phases $\rho_{a,b}$. The variable optical attenuators (VOA) set the average output intensity of the pulses to bright (classical regime) or dim (quantum regime).
\textbf{c}, Discretisation of the phase space to identify the twin fields during the public discussion.
}
\label{fig_1}
\end{figure*}

To introduce the new scheme, we plot in Fig.~\ref{fig_1}a a number of conceptual bounds for the key rate-versus-distance dependence of QKD, under ideal experimental conditions (see parameters in the inset). Lines $a-d$ represent key rates of quantum schemes obtained without resorting to a quantum repeater\cite{BDC+98,DLC+01,SSD+11}, hence they are denoted ``repeater-less bounds''. Line $d$, in particular, is the secret key capacity (SKC) of an optical quantum channel with losses\cite{PLO+17}, which quantifies the maximum secret information that can be transmitted in QKD\cite{TGW14}. On the experimental side, the key rates currently achieved are represented by the red symbols. They show a similar dependence on distance to the repeater-less bounds, but with lower rates, due to source and detector losses and other experimental imperfections. This highlights a limitation of existing QKD schemes, i.e., they can never surpass the SKC bound.

With the aid of a quantum repeater\cite{BDC+98,DLC+01,SSD+11}, it would be possible to overcome this barrier. Despite recent advances\cite{JTN+09,MSD+12,ATL15,ATM15}, however, such a device still remains difficult to realise. One of the simplest versions, tailored for intercity distances\cite{ATM15}, avoids using quantum memories and quantum error correction, but still requires non-demolition measurements, conditional optical switches and the multiplexing of a large number of single photon sources, all of which is far from trivial to implement. As a result, there is yet to be an experimental realisation of a scheme that surpasses the SKC barrier.
It is worth mentioning that although a trusted-node network\cite{Qiu14} and the use of satellites\cite{YCL+17} can greatly extend the reach of quantum communications, they do not exceed the SKC barrier. In the former case, the information ceases to be quantum at each intermediate node. For the latter, outer space provides a low-loss propagation medium, but the key rate per loss unit remains unchanged.

On the other hand, the scheme presented here can overcome the point-to-point SKC\cite{PLO+17}. We anticipate the TF-QKD key rates in Fig.~\ref{fig_1}a (thick lines). As shown in the figure, the ideal TF-QKD (dashed line) overcomes the repeater-less bounds after 200 km of standard optical fibre (lighter pink-shaded area). Even when realistic parameters are considered (solid line), TF-QKD can surpass the ideal repeater-less bound after 340~km of optical fibre (darker pink-shaded area). The gradient of the TF-QKD rates resembles that of a single quantum repeater connecting two end points\cite{Pir16}, also plotted in the figure. While conventional QKD's key rate scales linearly with the channel transmittance $\eta$ when $\eta\ll1$, TF-QKD's one scales with $\sim\eta^{1/2}$, thus dramatically improving the key rate-distance figure. Although a rigorous proof of unconditional security is beyond the scope of the current paper, this change in the loss dependence constitutes a fundamental advance in the long standing QKD paradigm.

In TF-QKD, dim optical pulses are generated by two light sources, phase-encoded with secret bits and sent to interfere\cite{PM67} on the beam splitter of an intermediate station, Charlie, who can even be a malicious party. Depending on which detector clicks, Charlie can infer whether the secret bits of the users (Alice and Bob) are equal (00 or 11) or different (01 or 10), but he cannot learn their absolute values (0 or 1).
This feature guards the scheme against eavesdropping, in a similar manner to phase-based measurement-device-independent (MDI)-QKD\cite{TLF+12,BM16}. { However, TF-QKD also employs phase randomisation and decoy states\cite{Hwa03,Wan05,LMC05} to considerably extend the distance of secure quantum communications. This, in turn, resembles decoy-state MDI-QKD\cite{LCQ12}}. There, the users send two photons, one each, to the central station to cause a two-photon interference followed by a coincidence count in Charlie's detectors. In TF-QKD, on the other hand, they send two optical fields, to produce a single-photon interference followed by a single-photon detection event. This lets TF-QKD retain the MDI characteristic, while gaining the square-root dependance of the key rate on the channel transmittance. Moreover, this provides an advantage over MDI-QKD even at short distances when Charlie's detectors have low efficiency.

As depicted in Fig.~\ref{fig_1}b, TF-QKD adopts the same components as decoy-state MDI-QKD, hence it can be implemented readily. However, it requires the coordinated phase randomisation of the twin fields. This is initially performed by Alice and Bob independently of each other, picking values $\rho_{a}$ (Alice) and $\rho_{b}$ (Bob) at random in the semi-open interval $[0,2\pi)$, similar to what has been suggested for the error correction routine of MDI-QKD\cite{MR12}. The phase interval is split into $M$ phase slices $\Delta_k=2\pi k/M$, $k=\{0,...,M-1\}$ (see example in Fig.~\ref{fig_1}c) from which partial phase slices $\Delta_{k(a)}$ and $\Delta_{k(b)}$ are defined for Alice and Bob, respectively.
The phase values randomly picked by the users necessarily fall in one of the phase slices. To identify the twin fields, the users publicly reveal $\Delta_{k(a,b)}$ together with the preparation bases. They keep only the runs with matching values and discard all the others. This entails that $\rho_{a}$ and $\rho_{b}$ will always differ by less than $2\pi /M$ for a pair of twin fields and there will be an intrinsic quantum bit error rate (QBER) $E_M$ due to the twins being close but not exactly identical. On average, it will be
\begin{equation}
E_M=\frac{M}{2\pi}\int_0^{2\pi/M} dt \sin^2 \frac{t}{2} = \frac{1}{2} - \frac{\sin (2\pi/M)}{4\pi/M}.
\label{eq:epr}
\end{equation}
This QBER tends to zero for $M\rightarrow\infty$. However, the probability of matching two phase slices scales with $1/M$. As a consequence there exists an optimal $M$ that guarantees the best performance. We run a realistic simulation to maximise the darker pink-shaded area in Fig.~\ref{fig_1}a and obtained the optimal value $M_{\textrm{opt}} = 16$, in correspondence of which $E_{M_{\textrm{opt}}}=1.275\%$.

\begin{figure*}
\centering
\includegraphics[width=0.75\textwidth]{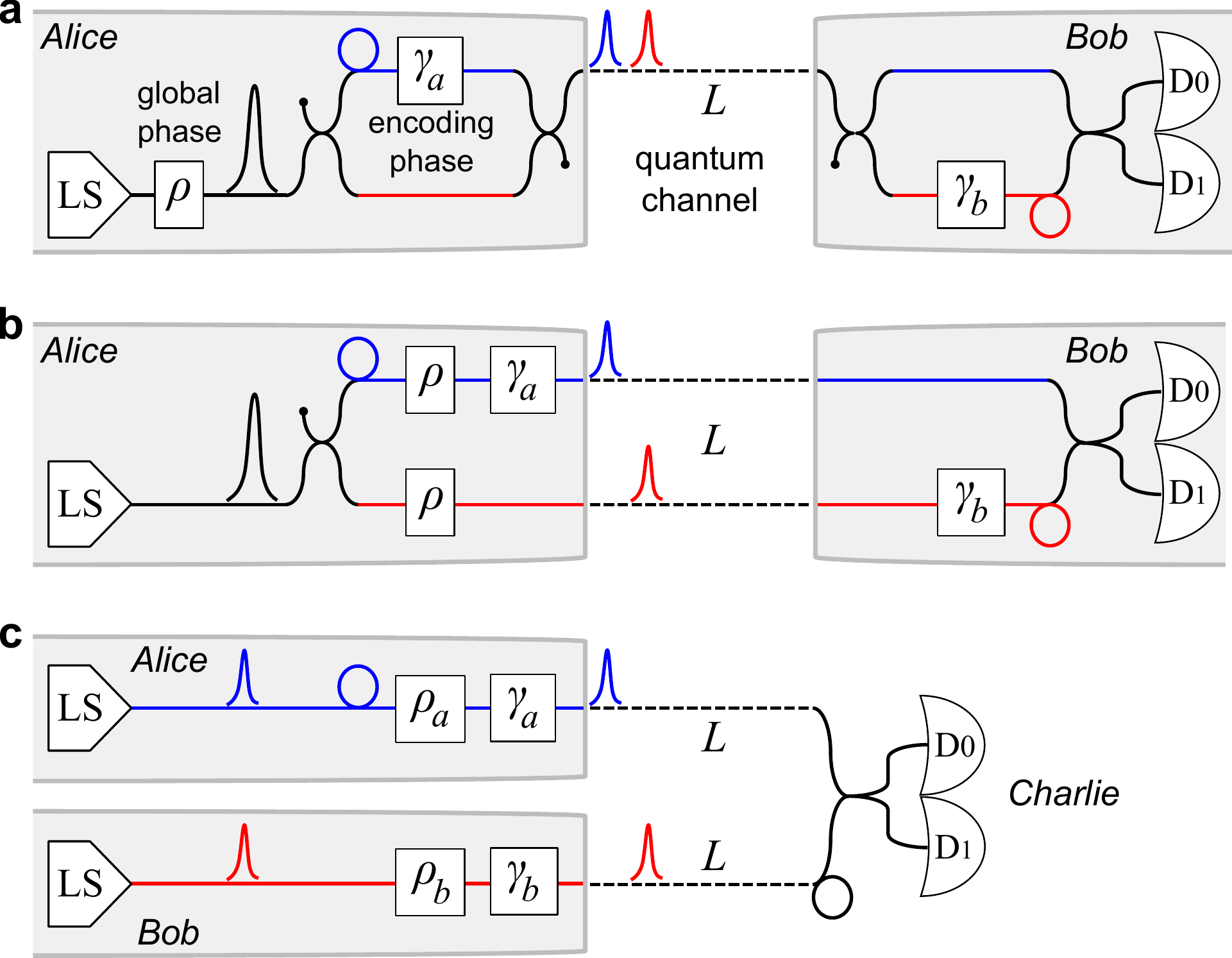}
\caption{\textbf{Diagrams for the quantum distribution of encryption keys}. The grey-shaded areas are inaccessible to the eavesdropper.
\textbf{a}, Typical phase-based QKD setup. A light source (LS) emits optical pulses with random global phase $\rho$. The primary pulse is split in two sub-pulses at the input of an asymmetric Mach-Zehnder interferometer (AMZI). The pulse on the longer path (blue) acquires an additional phase $\gamma_a$ respect to the other pulse (red). The pulses are sent on a quantum channel of length $L$ towards the receiving user (Bob), who owns a matched AMZI.
\textbf{b}, Unfolded QKD setup. The common path of length $L$ in Fig.~\ref{fig_2}a is now split in two separate paths of equal length $L$. The two secondary pulses travel on separate quantum channels to then interfere on Bob's beam splitter and eventually be detected.
\textbf{c}, Scheme analysed in this work. Alice and Bob are both transmitters. Each of them is provided with one laser source and one interferometer arm. Alice (Bob) prepares an optical pulse with random phase $\rho_a$ ($\rho_b$) and encoding phase $\gamma_a$ ($\gamma_b$) and transmits it on the quantum channel. Charlie overlaps the input pulses on the beam splitter and measures them. After he announces which detector clicked, the users reveal the basis values in $\gamma_{a,b}$ and the phase slices containing $\rho_{a,b}$.
}
\label{fig_2}
\end{figure*}

In Fig.~\ref{fig_2} we relate the new scheme to conventional QKD. We first represent the typical interferometer for a phase-encoded QKD setup (Fig.~\ref{fig_2}a). The light source generates a coherent state $\ket{e^{i\rho}\sqrt{\mu}}$, with $\mu$ the intensity and $\rho$ the electromagnetic phase carrying the so-called ``global phase information''. The phase $\rho$ is uniformly random and the actual state averaged over repeated runs is
$
\int_0^{2\pi} \frac{d\rho}{2\pi} \ket{ e^{i\rho} \sqrt{\mu} } \bra{ e^{i\rho} \sqrt{\mu} } = \sum_{n=0}^{\infty} p_{n|\mu} \ket{n} \bra{n},
$
where $p_{n|\mu}=e^{-\mu}\mu^n/n!$ is the (Poisson) probability to emit $n$ photons when a state with intensity $\mu$ was prepared. When the tagging argument\cite{GLLP} is applied to the efficient BB84 protocol\cite{LCA05} endowed with decoy states, the QKD key rate in the asymptotic scenario is given by\cite{LMC05}
\begin{equation}
R_{\textrm{QKD}}{ (\mu, L)} = \left. { \underline{Q}_1} \right|_{\mu,L} \left[1-h\left(\left.\overline{e}_1\right|_{\mu,L} \right)\right] - f Q_{\mu,L} h(E_{\mu,L}).
\label{QKD-rate}
\end{equation}
In Eq.~\eqref{QKD-rate}, we have explicitly written, for later convenience, the dependence on the total intensity $\mu$ and on the distance $L$ between Alice and Bob.
$\underline{Q}_1 = p_{1|\mu} \underline{y}_1$ is the lower bound for the single-photon gain; $\underline{y}_1$ and $\overline{e}_1$ are, respectively, the lower bound for the single-photon yield and the upper bound for the single-photon phase error rate, estimated through the decoy-state technique; $Q$ and $E$ are the gain and the QBER measured in the QKD session; $f$ accounts for the efficiency of error correction and $h$ is the binary entropy.

As an intermediate step to the new scheme, the QKD interferometer of Fig.~\ref{fig_2}a has been unfolded in Fig.~\ref{fig_2}b, where the two pulses travel now on separate channels and are separately encoded with the same phase $\rho$. These are the twin fields that will interfere on Charlie's beam splitter. The emitted state is unchanged from the previous scheme, as is the disclosed classical information, so the two schemes are equivalent.

In Fig.~\ref{fig_2}c we present the TF-QKD scheme. The detectors have been outsourced to Charlie and the users' stations have been separated, so that Bob's station is now located at distance $2L$ from Alice. The users' lasers emit optical pulses that interfere\cite{PM67} on Charlie's beam splitter. The pulses are encoded with random phases $\rho_{a,b}$, which will then be revealed to a finite precision through the public announcement of the phase slices $\Delta_{k(a,b)}$.
We notice that this is different from conventional QKD, where the value of the global phase is never revealed.

The key feature of TF-QKD is the doubling of the distance between Alice and Bob. As it can be seen from Fig.~\ref{fig_2}, the red and blue pulses travel a distance $L$ each, both in QKD and in TF-QKD. However, while in QKD they co-propagate from Alice to Bob, in TF-QKD they run from Alice and Bob towards Charlie, thus effectively increasing the transmission distance.

In the SI, we show that if revealing the global phase $\rho$ after Charlie's measurement did not contribute to the eavesdropper's information, the TF-QKD key rate could be expressed through Eq.~\eqref{QKD-rate}, as
\begin{equation}
 R_{\textrm{TF-QKD}}^{ (\lnot\rho)}(\mu, L) = \frac{d}{M} \[ R_{\textrm{QKD}}\(\mu, \frac{L}{2}\) \]_{\oplus E_{M}}.
\label{TF-QKD-rate}
\end{equation}
However, the public disclosure of $\rho$, even after Charlie's measurement, can leak information to the eavesdropper (Eve). In the SI, we consider a specific attack built on this leakage and show that the resulting key rate still overcomes the SKC at long distance. Despite that, we stress that Eq.~(3) does not cover the most general attack by Eve and that the analysis of general attacks is an outstanding challenge.

The notation $\oplus E_{M}$ in Eq.~\eqref{TF-QKD-rate} prompts the intrinsic QBER of TF-QKD, $E_M$, due to its phase-randomisation. The total intensity of the optical pulses is $\mu=\mu_a+\mu_b$, with $\mu_a$ ($\mu_b$) the intensity of the pulses emitted by Alice (Bob). The coefficient $1/M$ stems from sifting the phase slices whereas $d$ is the duty cycle between the classical and the quantum modalities, described later on. Eq.~\eqref{TF-QKD-rate} makes it apparent that a distance $L/2$ in QKD corresponds to a distance $L$ in TF-QKD.


The main technical challenge in implementing TF-QKD is controlling the phase evolution of the twin fields, which travel hundreds of kilometres before interfering on Charlie's beam splitter. 
The differential phase { fluctuation} between the two optical paths { linking the users to Charlie} can be written as
\begin{equation}
\delta_{ba}=\frac{2\pi}{s}(\Delta \nu L+\nu \Delta L),
\label{eq:phase-drift}
\end{equation}
where $s$ is the speed of light in the fibre. The first term arises from the frequency difference $\Delta \nu$ of the users' lasers and can be easily compensated using phase-locking techniques\cite{SCL+94} routinely employed in optical communications\cite{AML09}. With { a feasible value} $\Delta \nu <1$~Hz\cite{LPW16}, the phase uncertainty would be $\sim 0.01$~rad over 300~km of fibre, negligibly contributing to the QBER.
The second term represents a more serious impairment. During the propagation in the very long fibres, the twin fields travel different paths, so their relative phase will vary. The phase drift of a fibre-based Mach-Zehnder interferometer with 36.5~km-long arms was previously characterised to be around $0.3 - 1$ rad/ms\cite{MDS+08}.

\begin{figure*}
\centering
\includegraphics[width=0.75\textwidth]{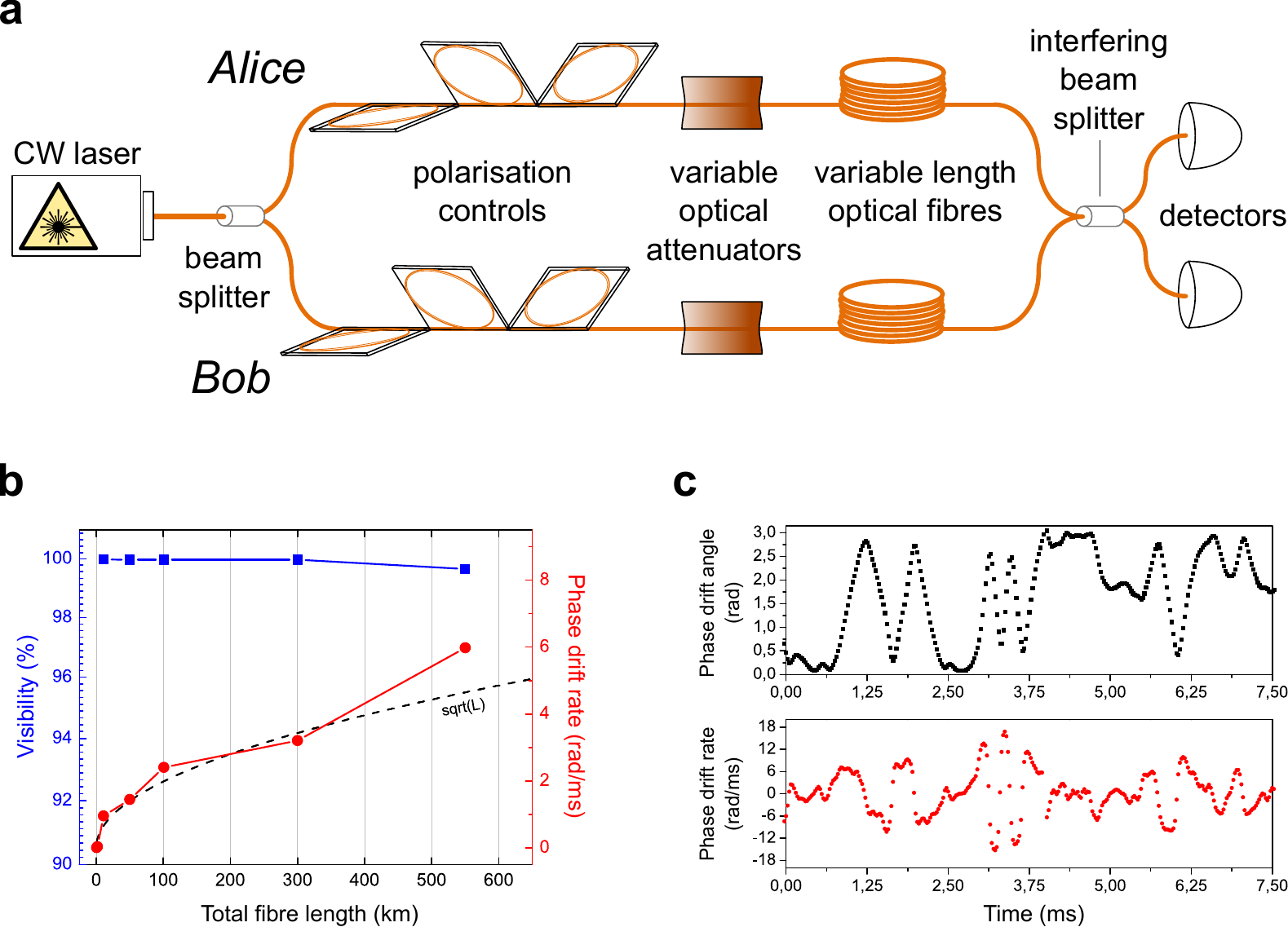}
\caption{\textbf{Experimental characterisation of phase drift and visibility.}
  \textbf{a}, Experimental setup. A light beam emitted by a continuous-wave (CW) laser is sent through the two arms of the interferometer. Polarization controls are used to set the correct polarisation, which remains stable for a time much longer than the phase drift scale. Variable optical attenuators equalise the intensity of the fields entering the interfering beam splitter. Two equal reels of single-mode optical fibre connect the preparation stage to the beam splitter and the detection stage, where a power meter (Keysight 7748A) with a sampling rate of 40~kHz and power range between $-110$~dBm and $10$~dBm is used to monitor the phase drift.
  \textbf{b}, Maximum visibility obtained in the experiment (blue) and phase drift rate (red) as function of distance. The dashed line represents a qualitative fit that assumes a random walk model for the phase drift.
  \textbf{c}, Measured phase drift (top, black) and related phase drift rate (bottom, red) in the longest-distance configuration of 550~km, obtained with two fibre spools of length 275~km each. The maximum visibility observed at this specific distance is 99.65\%.
}
\label{fig_3}
\end{figure*}

To determine the phase drift over much longer fibres, we used the experimental setup shown in Fig.~\ref{fig_3}a. The presence of a single laser assures that $\Delta \nu=0$ in Eq.~\eqref{eq:phase-drift}, thus letting us measure only the noise due to the fluctuations in the channel. The measured phase drift rate follows a Gaussian distribution with zero mean and standard deviation equal to 2.4~rad/ms at a total distance of 100~km and 6.0~rad/ms at the longest distance of 550~km (Figs.~\ref{fig_3}b and \ref{fig_3}c). Compensating the phase drift would require bright pulses and active feedback, to be realised by Charlie acting on his phase modulator (details in SI).
Fig.~\ref{fig_3}b also shows the visibility measured as a function of the fibre length. The visibility remains $>99.65\%$ for all distances, thus causing a negligible $0.175\%$ contribution to the QBER due to a loss of coherence along the fibre.

Our findings suggest that the point-to-point secret key capacity of a quantum channel can be overcome without using quantum repeaters, with a scheme that borrows components and techniques from ordinary QKD. This is not at variance with existing results\cite{TGW14,PLO+17}, as TF-QKD is not point-to-point. Moreover, it exploits an assumption that is not present in the secret key capacity bounds. As in MDI-QKD, the security of TF-QKD does not depend on the measurement devices. At the same time, its single-photon nature entails count and error rates similar to standard QKD. Further work is necessary to prove the unconditional security of TF-QKD, which is an important open question left for future investigation. We expect that the counter-intuitive features of the new scheme will stimulate further research extending the limits of QKD.


\noindent\textbf{Supplementary Information} is attached.\\

\noindent\textbf{Acknowledgements} We gratefully acknowledge Kiyoshi Tamaki for his constructive criticism on the security argument. We acknowledge useful discussions with X.~Ma, N. L\"{u}tkenhaus, B. Fr\"{o}hlich, R.~M.~Stevenson, D.~Marangon and A.~J.~Bennett.\\

\noindent\textbf{Author contributions} M.L. and Z.L.Y. developed the TF-QKD scheme. Z.L.Y. and J.F.D. setup and performed the experiments and all authors analysed the results. A.J.S. guided the work. M.L. wrote the manuscript with contributions from all the authors.\\

\noindent\textbf{Author Information} The authors declare no competing interests.\\



\end{document}